\documentclass{PoS}

\title{Constraints on the cosmic ray cluster physics from a very deep observation of the Perseus cluster with MAGIC}

\ShortTitle{Perseus cluster observation with MAGIC: CR physics}

\author{\speaker{Pierre Colin}\\%\thanks{A footnote may follow.}\\
        Max-Planck-Institut f\"ur Physics, Munich, Germany \\
        E-mail: \email{colin@mppmu.mpg.de}}

\author{Fabio Zandanel\\
         GRAPPA Institute, University of Amsterdam, The Netherlands \\
        E-mail: \email{f.zandanel@uva.nl}}

\author{Monica Vazquez Acosta$^{a}$, Joaquim Palacio$^{b}$ for The MAGIC collaboration,\\
  $^{a}$Instituto de Astrofisica de Canaria, La Laguna (Tenerife), Spain \\
  $^{b}$Institut de Fisica d'Altes Energies, Bellaterra (Barcelona), Spain
  }
  
\author{Christoph Pfrommer$^{c}$ and Anders Pinzke$^{d}$\\
  $^{c}$HITS, Schloss-Wolfsbrunnenweg 33, Heidelberg, Germany\\
  $^{d}$Dark Cosmology Center, University of Copenhagen, Denmark\\
  }

\abstract{Galaxy clusters are the largest and most massive gravitationally bound structures known in the Universe.
Cosmic-Ray (CR) hadrons accelerated at structure formation shocks and injected by galaxies, are confined in galaxy clusters
where they accumulate for cosmological times. The presence of diffuse synchrotron radio emission in several clusters proves
the existence of high-energy electrons, and magnetic fields. However, a direct proof of CR proton acceleration is missing.
The presence of CR protons can be probe through the diffuse $\gamma$-ray emission induced by their hadronic interaction
with the Intra-Cluster Medium (ICM).
The Perseus cluster, a nearby cool-core cluster, has been identified to be among the best candidates to detect such emission.
We present here the results of a very deep observation of the Perseus cluster with the MAGIC telescopes,
accumulating about 250 hours of data from 2009 to 2014. No evidence of large-scale very-high-energy $\gamma$-ray emission
from CR-ICM interactions has been detected. 
The derived flux upper limits in the TeV regime allow us to put stringent constraints 
on the physics of cluster CRs, in particular on the
CR-to-thermal pressure, the CR acceleration efficiency at formation shocks and the
magnetic field of the central cluster region.}

\FullConference{The 34th International Cosmic Ray Conference,\\
		30 July- 6 August, 2015\\
		The Hague, The Netherlands}

\begin{document}

\section{Introduction}

Clusters of galaxies represent the latest stage of the structure formation, 
produced by merging small group of galaxies and gas accretion.
%They represent a sort of microcosm example of the Universe itself, and
Their study is a powerful cosmological tools to test the evolution model of the Universe \cite{2005RvMP...77..207V}.
Cosmic-ray (CR) protons can be accelerated by structure formation shocks, and outflows from galaxies and active galactic nuclei (AGNs) of
the cluster, and accumulate in it for cosmological times.
These CR protons can interact hadronically with the protons of the intra-cluster medium (ICM),
a hot thermal plasma with $k_BT\sim$keV, and generate pions.
While $\pi^\pm$'s decay to secondary electrons and neutrinos, the $\pi^0$'s decay directly to high-energy $\gamma$ rays.
Despite many observational efforts, diffuse $\gamma$-ray emission from clusters has remained undetected.
%\footnote{For space-based cluster observations in the GeV-band, see 
%\cite{2003ApJ...588..155R, 2010JCAP...05..025A,
%2010ApJ...717L..71A,2011ApJ...728...53J,2012arXiv1207.6749H,2012JCAP...07..017A,
%2013arXiv1308.6278H,2014MNRAS.440..663Z,
%2013arXiv1308.5654T,2013arXiv1309.0197P,2014arXiv1405.7047G,2015arXiv150502782V}. For ground-based 
%observations in the energy band 
%above $100$~GeV, see \cite{2006ApJ...644..148P, 
%2008AIPC.1085..569P,2009arXiv0907.0727T, 2009A&A...495...27A,
%2009arXiv0907.3001D,2009arXiv0907.5000G,cangaroo_clusters, 
%2009ApJ...706L.275A,2010ApJ...710..634A,2011arXiv1111.5544M,
%2012...VERITAS,2012A&A...545A.103H}.}

Non-thermal emission is observed at radio frequencies in many galaxy clusters
in the form of diffuse synchrotron radiation \cite{2012A&ARv..20...54F}.
This probes the presence of relativistic CR electrons and magnetic fields in the cluster environment.
But, a proof for CR-proton acceleration has yet to be found.
%The observed CR electrons can also produce hard X-rays by inverse-Compton (IC) scattering of cosmic 
%microwave background (CMB) photons. Several claims of IC detection have been made in the past (see 
%\citealp{2008SSRv..134...71R} for a review), but more recent observations do not confirm them 
%\citep{2009ApJ...690..367A,2010ApJ...725.1688A,2011ApJ...727..119W,
%2012ApJ...748...67W, 2014ApJ...792...48W,2015ApJ...800..139G}, and the possible diffuse IC emission from 
%clusters remains elusive too.
The observed diffuse radio emission in clusters can be divided in three main 
categories: peripheral radio relics, central radio halos and mini-halos \cite{2012A&ARv..20...54F,2014IJMPD..2330007B}.
%The latter are usually divided in two further categories: giant-halos hosted in merging, 
%non-cool-core clusters (e.g., the Coma cluster  
%\cite{1997A&A...321...55D,2011MNRAS.412....2B}), and mini-halos hosted in 
%relaxed, cool-core clusters (e.g., the Perseus cluster \cite{1990MNRAS.246..477P,Sijbring1993}).
While radio relics are though to be related to merger shocks, the origin of radio halos and mini-halos 
have been historically debated between re-acceleration
%\footnote{See, e.g., 
%\cite{1987A&A...182...21S,2001MNRAS.320..365B,2001ApJ...557..560P,
%2002A&A...386..456G,2002ApJ...577..658O,2003ApJ...584..190F,2004MNRAS.350.1174B,
%2005MNRAS.363.1173B,2005MNRAS.357.1313C,2007MNRAS.378..245B,2010arXiv1008.0184B,
%2012arXiv1207.3025B,2013MNRAS.429.3564D,2013ApJ...762...78Z,2015arXiv150307870P} .} 
and hadronic models.
%\footnote{See, e.g., 
%\cite{1980ApJ...239L..93D,1982AJ.....87.1266V,1997ApJ...477..560E,
%1999APh....12..169B,2000A&A...362..151D, 2001ApJ...559...59M, 
%2003MNRAS.342.1009M,2003A&A...407L..73P,2003APh....19..679G,2004A&A...413...17P, 
%2004MNRAS.352...76P,2007IJMPA..22..681B, 2008MNRAS.385.1211P, 
%2008MNRAS.385.1242P,2009JCAP...09..024K,2010MNRAS.401...47D, 
%2010arXiv1003.0336D,2010arXiv1003.1133K,2010MNRAS.409..449P,2011arXiv1105.3240P,
%2011A&A...527A..99E,2012ApJ...746...53F,2012arXiv1207.6410Z}.}
In the re-acceleration model, a seed population of CR electrons is re-accelerated 
by turbulence, while in the hadronic 
scenario the radio-emitting electrons are secondaries produced by CR protons 
interacting with the ICM. 
%Currently, the re-acceleration scenario is favoured for 
%giant radio halos, while for mini-halos both the re-acceleration and hadronic 
%models can explain the observed emission
%(see, e.g., \cite{2011A&A...527A..99E} and \cite{2014IJMPD..2330007B} for extensive discussions).
Actually, the presence and role of CR protons in clusters can be probed directly only
through the $\gamma$-rays and neutrinos induced by hadronic interactions.  
The high-energy astronomy window is then crucial to understand non-thermal phenomena in galaxy clusters. 
%(this is also true, albeit even more challenging, for neutrinos; see, e.g, \cite{2014arXiv1410.8697Z}).

The Perseus cluster of galaxies (Abell~426) is a relaxed, cool-core 
cluster located at a distance of about 78\,Mpc ($z$=0.018).
It hosts the brightest thermal X-ray emission from the ICM \cite{2003ApJ...590..225C} and a very luminous radio mini-halo 
\cite{1990MNRAS.246..477P,Sijbring1993}. The high ICM density at the centre implies a high 
target density for CR hadronic interactions.
Therefore, Perseus is the best cluster where to search for CR-induced $\gamma$-ray emission 
(see \cite{2010MNRAS.409..449P,2010ApJ...710..634A,2011arXiv1105.3240P,2011arXiv1111.5544M} for a detailed argumentation).
This cluster also hosts two $\gamma$-ray bright AGNs: NGC\,1275, the central dominant galaxy of the cluster,    
%NGC\,1265, archetype of head-tail radio galaxy in which the jets are bent by their interaction with the ICM,
and IC\,310, a peculiar object that could be an intermediate state between a BL Lac and a radio galaxy.
Both were detected with \emph{Fermi}-LAT \cite{2009arXiv0904.1904T,2010A&A...519L...6N}
and MAGIC \cite{2010ApJ...723L.207A,2012A&A...539L...2A}.
%NGC\,1275 obscures the central diffuse cluster emission over most of the $\gamma$-ray band,
%in particular below the spectral break around 10\,GeV \cite{2014A&A...564A...5A}.

Since 2008, the Perseus cluster is intensively observed by the MAGIC telescopes. Here, we present 
the results obtained with $\sim$250~hours of effective observation time taken in stereoscopic mode from 2009 to 2014,
where constraints on the CR population in the cluster are derived.
%After describing the observation and data analysis in 
%Section~\ref{sec:2}, we report our observational results and searches for 
%CR-induced signal in Section~\ref{sec:3} and Section~\ref{sec:4}, respectively. 
%In Section~\ref{sec:5}, we discuss the interpretation of our constraints on the 
%CR physics in galaxy clusters, and, finally, in Section~\ref{sec:6} we present 
%our conclusions. Throughout the paper, we assume a standard $\Lambda$CDM 
%cosmology with $H = 70$\,km\,s$^{-1}$\,Mpc$^{-1}$.

\begin{figure*}[ht!]
\centering
\includegraphics[width=0.45\textwidth]{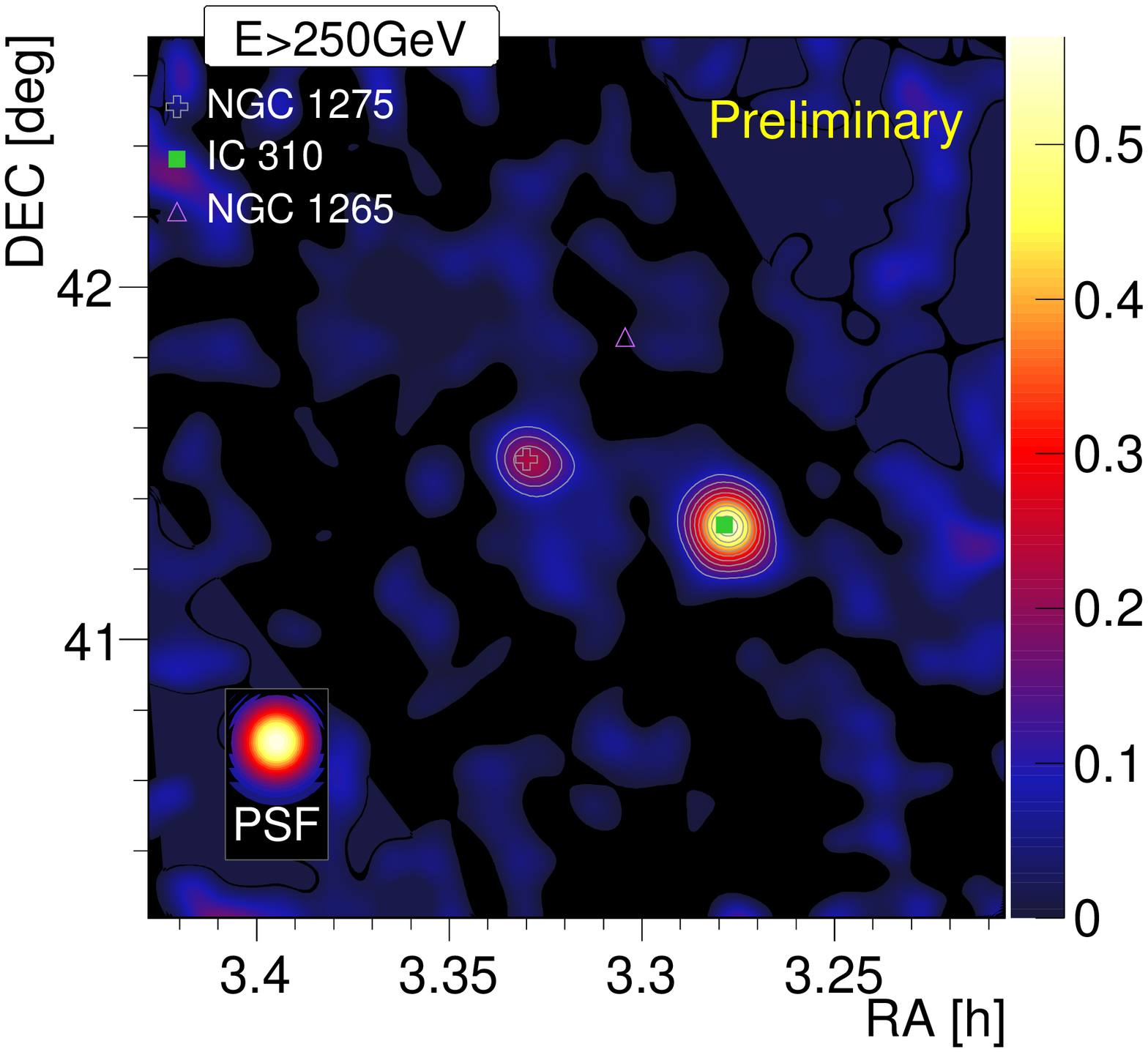}
\includegraphics[width=0.45\textwidth]{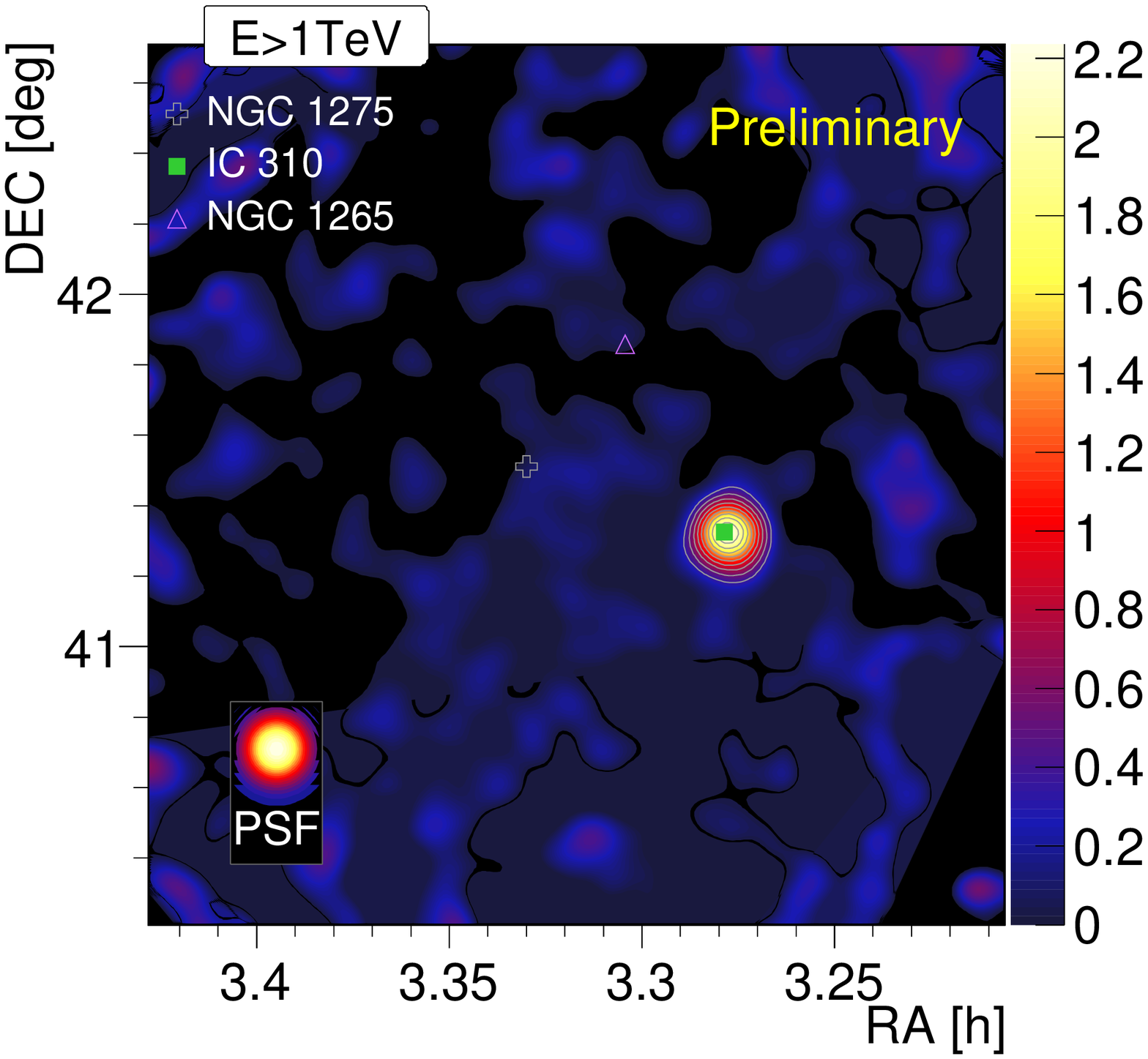}
\includegraphics[width=0.45\textwidth]{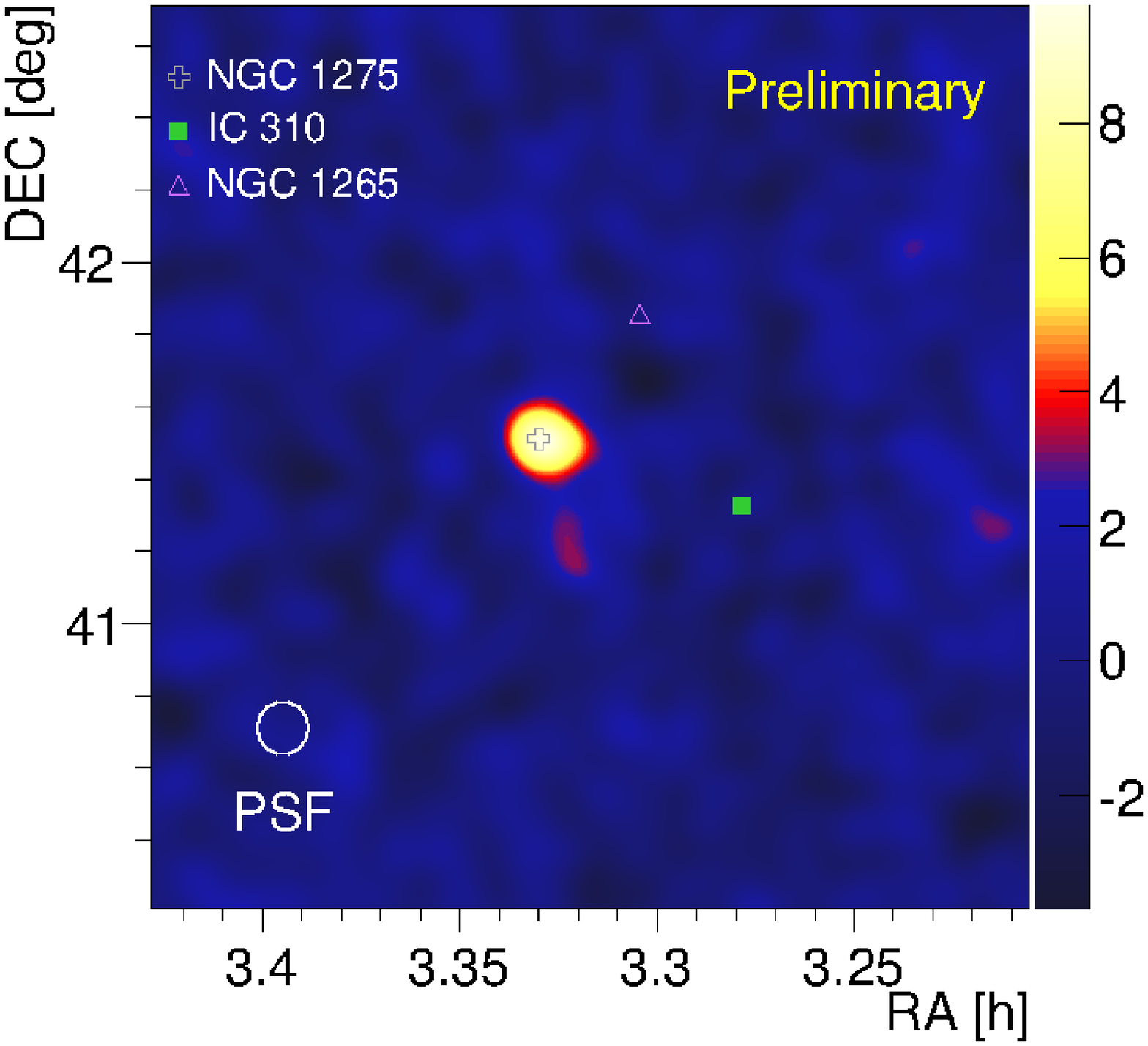}
\includegraphics[width=0.45\textwidth]{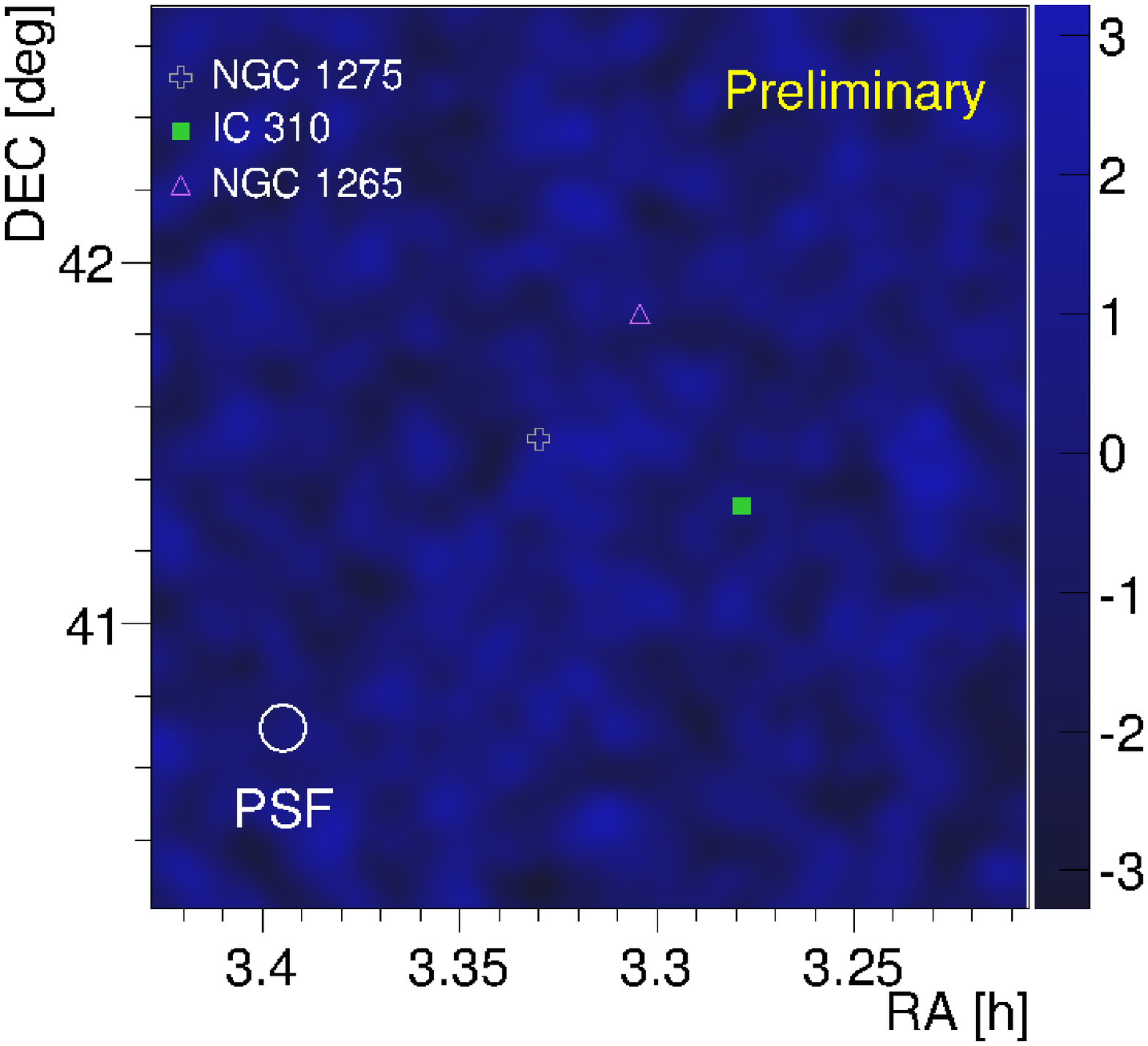}
\caption{Perseus cluster sky maps for an energy threshold of 250\,GeV (left-hand 
panels) and 1\,TeV (right-hand panels) obtained from $\sim$250\,h of MAGIC 
observation. Top panels show the relative flux (colour code, expressed in 
signal-to-background ratio) and the excess significance (contour lines starting 
from 4$\sigma$ with steps of 2$\sigma$). Lower panels show the significance 
maps where the signal from IC\,310 is included in the background model. Symbols 
indicate the positions of the three brightest radio galaxies of the cluster.}
\label{fig:SkyMaps}
\end{figure*}

\section{MAGIC observations and results}
\label{sec:2}
MAGIC is a system of two 17\,m diameter Imaging Atmospheric Cherenkov Telescopes 
located on the Canary island of La Palma. It observes the $\gamma$-ray sky 
from $\sim$50\,GeV to more than 50\,TeV. 
The observation of the Perseus cluster started in 2008 with $\sim$24\,h of 
observation with a single telescope, that did not allow any detection \cite{2010ApJ...710..634A}.
Since 2009, MAGIC operates in stereoscopic mode, providing a much better sensitivity \cite{2014arXiv1409.5594}. 
%The stereo observations can be separated in two periods, before and after a 
%major upgrade of the telescopes occurring during the summers 2011 and 2012 \cite{2014arXiv1409.6073}.
Here, we combine all the Perseus-cluster stereo observations carried out from October 2009 to November 2014.
%performed during these two periods.
The observations of a first period, 2009--2011, taken solely during dark time at low zenith angle
(from 12$^\circ$ to 36$^\circ$), led to the discovery of IC\,310 \cite{2010ApJ...723L.207A} and 
NGC\,1275 \cite{2012A&A...539L...2A}.
The observations of the period 2012-2014 were carried out after an upgrade of the telescopes \cite{2014arXiv1409.6073}.
They were taken with relaxed criteria including Moonlight-night and a larger zenith-angle coverage from 12$^\circ$ to 60$^\circ$.
After data-quality selection, the final sample in the two periods considered
amounted to an effective time of 85\,h and 168\,h, respectively.

The standard MAGIC Analysis and Reconstruction Software (MARS) is used to analyze the data.  
For the 2009--2011 data sample, we use the same analysis than in our previous study on 
cluster CR-physics \cite{2011arXiv1111.5544M}. For the 2012--2014 data sample, to properly handle the data taken under Moonlight
we use 33\% higher image cleaning levels than the standard levels reported in \cite{2014arXiv1409.5594}.
This increases the energy threshold but does not affect the performance at high energy.
%Additionally, only events with image size above 150 photo-electrons in both telescopes are kept, excluding events near the trigger threshold,
%where the instrument response is unstable during Moon time.

%The $\gamma$-hadron separation is performed by the standard MAGIC method using 
%Random Forest. The remaining background level induced by the possible CR-induced 
%flux, is estimated from mirror regions (OFF regions) at 0.4$^\circ$ from the 
%camera centre. To prevent contamination by the strong IC\,310 signal, the OFF 
%regions closer than 0.4$^\circ$ to IC\,310 have been excluded.
%For point-like source analysis, an average of 5 OFF regions in the field of view are used. In 
%the case of extended source analysis only the most distant OFF region, lying 
%0.8$^\circ$ from NGC\,1275, is used. 

%\section{MAGIC results}
%\label{sec:3}

The 253\,h of stereo observation, combined at the latest stage of the analysis, provide the deepest view of the Perseus cluster at 
very high energy. The right- and left-upper panels of Figure~\ref{fig:SkyMaps} 
show the relative-flux (i.e., signal-to-background ratio) sky maps for an energy 
threshold of 250\,GeV and 1\,TeV, respectively. A clear signal is detected from 
the two previously discovered AGNs.
The bright and hard source IC\,310 is visible in both maps with high 
significance and could mask smaller signals. In order to search for weak emissions, we included the point-like emission
from IC\,310 in our background model. The lower panels of Figure~\ref{fig:SkyMaps} show the resulting significance sky maps.
Above 250\,GeV, NGC\,1275 is detected with a significance $>$8$\sigma$. %and a signal-to-background ratio $>$20\%.
The shape of the detected signal is in perfect agreement with a point-like object
such as expected for an AGN.
Above 1\,TeV, however, no source other than IC\,310 is detected. For both energies, there is no sign of 
diffuse $\gamma$-ray structures inside the cluster.
%Figure~\ref{fig:NGC1275_theta_plot} compares the excess ($\gamma$-ray) event 
%distribution above 250\,GeV as a function of the squared distance to NGC\,1275 
%($\theta^2$) with the MAGIC Point Spread Function (PSF) obtained from 
%contemporaneous Crab Nebula data. The shape of the detected signal is in perfect 
%agreement with a point-like object such as expected for an AGN.

\begin{figure}[t!]
\centering
\includegraphics[width=0.6\textwidth]{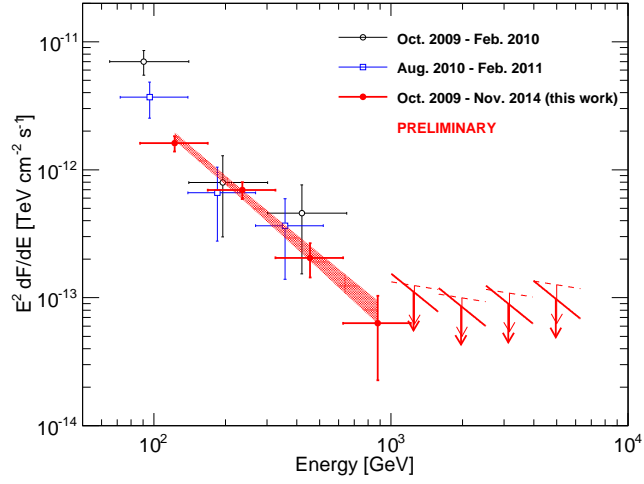}
\caption{Spectral energy distribution of NGC\,1275 averaged over different 
periods. The red symbols show our new result obtained with $\sim$250\,h of data.
The red arrows are the differential flux upper limits for a power-law spectrum with a photon index $\Gamma$=3.5 (thick 
solid lines) and $\Gamma$=2.3 (thin and dashed lines).}
\label{fig:NGC1275_sed}
\end{figure}

The average energy spectrum of NGC\,1275 obtained with the full 2009--2014 data set
is shown in Figure~\ref{fig:NGC1275_sed}, together with 
the previously reported spectra from the first two years of observation \cite{2014A&A...564A...5A}.
The spectrum of this analysis starts at higher energy because the data include Moonlight and 
large-zenith-angle observations. The statistical precision is largely improved, extending the spectrum
coverage with a data point around 880\,GeV.
%Note that this data point is only marginally significant ($\sim$2$\sigma$)
%and is in agreement with the upper limits discussed later.
The spectrum between 90\,GeV and 1200\,GeV can be described by a simple power law
%\footnote{Power-law fit obtained with the forward-unfolding method over 7 reconstructed-energy bins
%($\chi^{2}/n_{dof}=2.4/5$).}
\begin{equation}
\frac{\mbox{d}F}{\mbox{d}E} = f_{0} 
\left(\frac{E}{\mathrm{200\,GeV}}\right)^{-\Gamma},
\end{equation}
with a photon index $\Gamma=3.6\pm0.2_{stat}\pm0.2_{syst}$
and a normalization constant at 200\,GeV of $f_{0} =(2.1 \pm 0.2_{stat} \pm 
0.3_{syst}) \times 10^{-11} \mathrm{cm^{-2} s^{-1} TeV^{-1}}$. 

\section{Search for cosmic-ray induced emission}
\label{sec:4}
The $\gamma$-ray emission from the central galaxy NGC\,1275 is consistent with a point-like source
and no diffuse component is observed. The measured flux is much larger than what is 
expected from the CR-induced emission, and it could be outshining the latter.
However, the NGC\,1275 AGN spectrum is very steep and no signal is detected 
above about 1\,TeV, while models expect the CR-induced spectrum to be a flat power-law
with no cutoff in the MAGIC energy range \cite{2010MNRAS.409..449P}.
%(but at much higher energies, above the PeV, 
%due to high-energy CR protons no longer confined in the cluster volume; see, 
%e.g., \cite{1996SSRv...75..279V,1997ApJ...487..529B,2010MNRAS.409..449P}). 
Therefore, we use the energies above 1\,TeV to search for the possible
diffuse CR-induced component. 
%As the $\gamma$-ray emission above 10\,TeV is strongly reduced by its 
%interaction with the infrared Extragalactic Background Light (EBL), the best 
%energy range is between $\sim$1 and $\sim$10\,TeV. 

\begin{figure}[t!]
\includegraphics[width=0.5\textwidth]{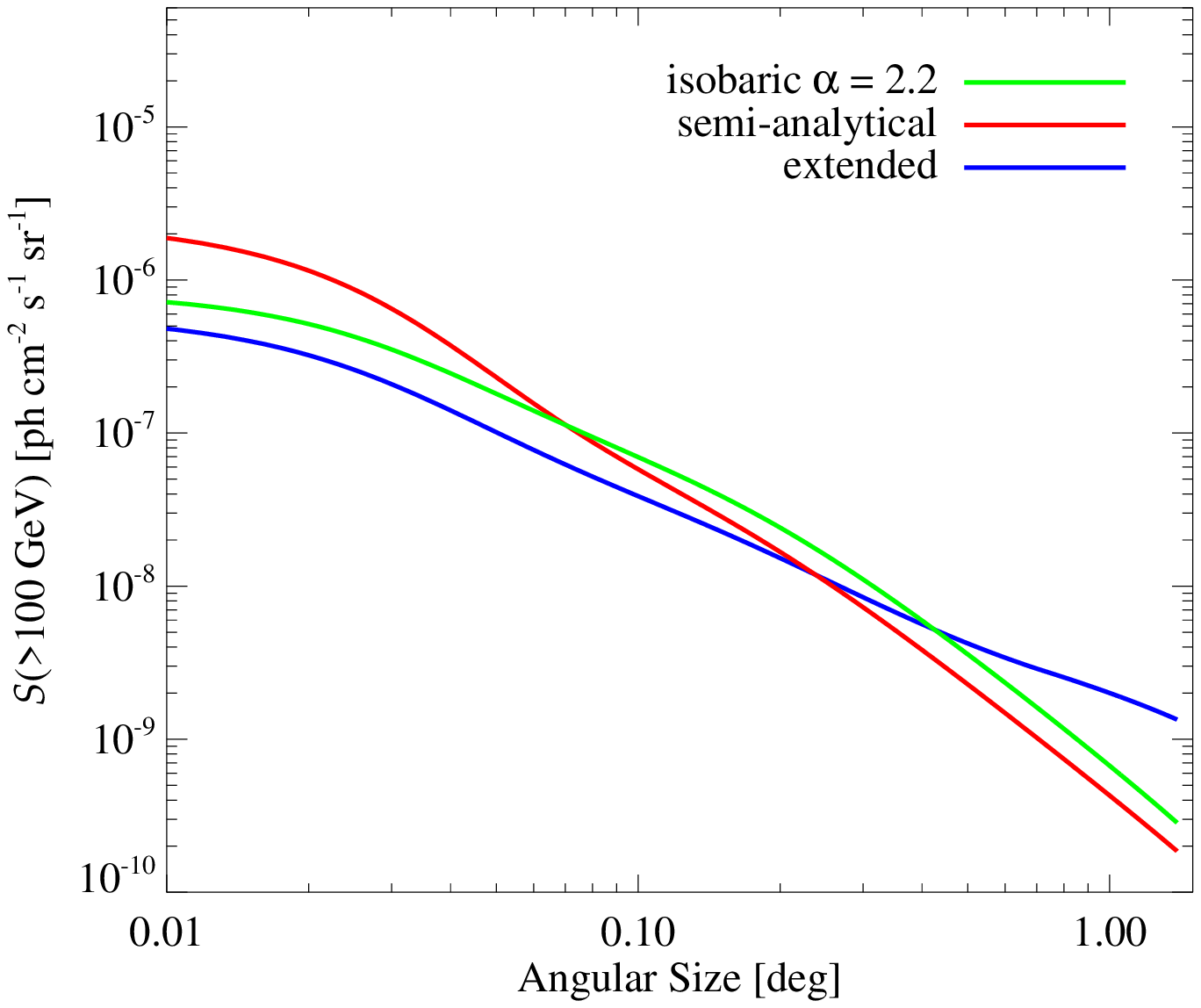}
\includegraphics[width=0.47\textwidth]{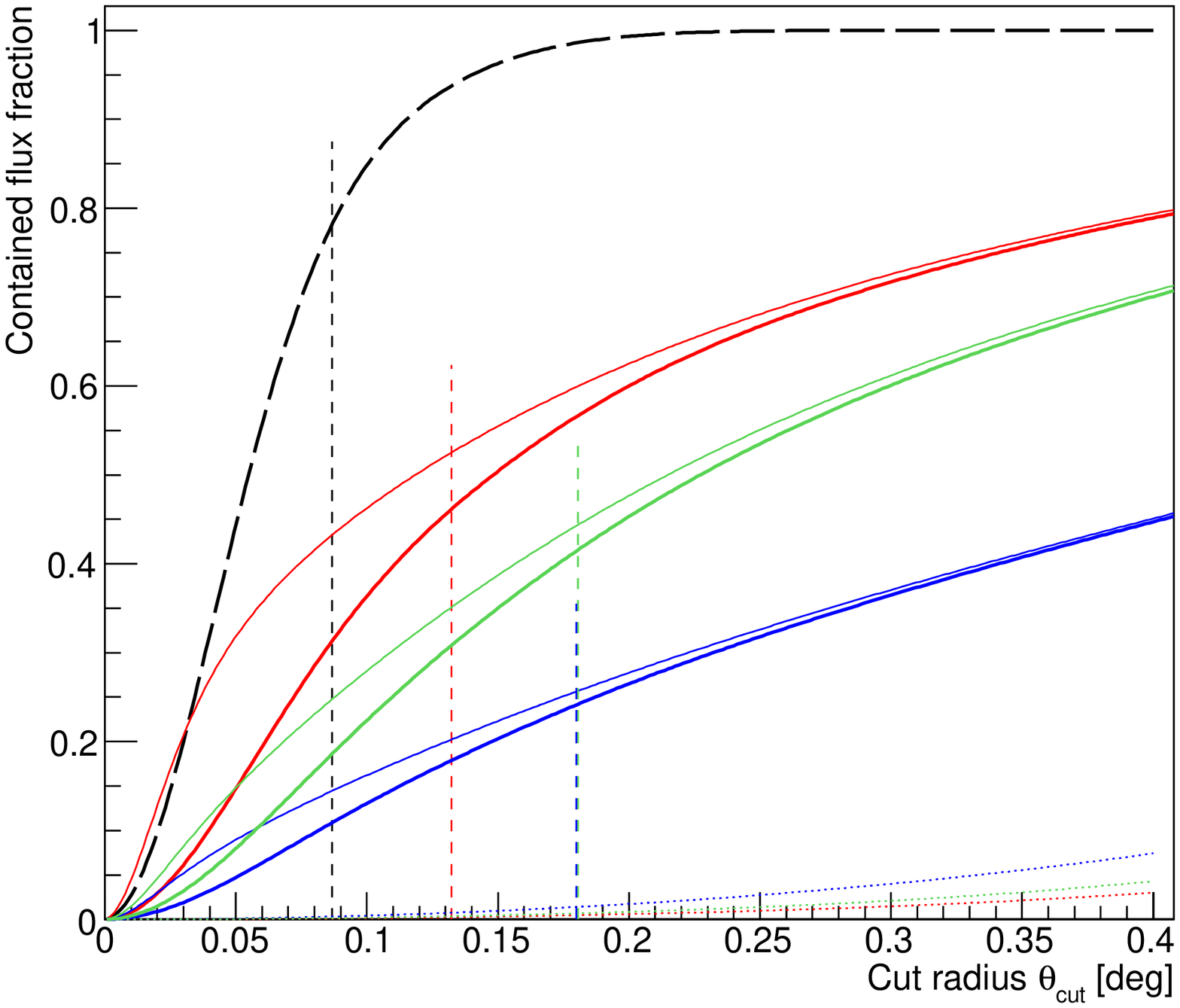}
\centering
\caption{\textbf{Left:} Surface brightness profiles of the three tested spatial templates
for the CR-induced emission in Perseus.
%\emph{isobaric} with $\alpha=2.2$ in green, \emph{semi-analytical} in red, and \emph{extended} in blue.
%The normalization of the \emph{isobaric} model is set to respect our 
%previous upper limits \cite{2011arXiv1111.5544M}, while the normalizations of 
%the \emph{semi-analytical} and \emph{extended} models are as from 
%\cite{2010MNRAS.409..449P} and \cite{2012arXiv1207.6410Z}, respectively.
%See main text for details.
\textbf{Right:} Cumulative fraction of signal within a given radius for different 
models: \emph{point-like} in black (long-dashed line), 
\emph{isobaric} in green, \emph{semi-analytical} in red, and \emph{extended} in blue.
The distribution of the \emph{point-like} model corresponds the the MAGIC PSF 
above 630\,GeV.
The thin and thick coloured solid lines represent the real and 
MAGIC-reconstructed, i.e., smeared by the PSF, signal fractions, respectively.
The dotted lines show the fraction of these signals contained in the background region.
The vertical dashed lines correspond to the used optimal $\theta_{cut}$ values.
}
\label{fig:TheoreticalProfiles}
\end{figure}

As the spatial shape of the CR-induced emission is not known, three
different models are used as template: i) the 
\emph{isobaric} model assuming a constant CR-to-thermal pressure $X_\mathrm{CR} 
= P_\mathrm{CR}/P_\mathrm{th}$ \cite{2004A&A...413...17P}, ii) the 
\emph{semi-analytical} model for CRs derived from hydrodynamical 
simulations of clusters \cite{2010MNRAS.409..449P}, and iii) the \emph{extended}
hadronic model of \cite{2012arXiv1207.6410Z}, where the possibility of CR 
propagation out of the cluster core is considered,
resulting in a significantly flatter CR profile. 
%\citep{2011A&A...527A..99E,2013arXiv1303.4746W}.
The $\gamma$-ray surface brightness of these three models is shown in Figure~\ref{fig:TheoreticalProfiles}.
%In all cases, fundamental input 
%parameters are the Perseus ICM density and temperature as measured in X-rays 
%\cite{2003ApJ...590..225C}. 
The fraction of signal expected within a circular region of a given radius 
$\theta$ from the cluster centre and the MAGIC Point Spread Function (PSF) above 630\,GeV are also shown. 
Since the predicted CR-induced signal extension is significantly larger than the 
PSF, optimal selection cuts $\theta_{cut}$, different than 
for a point-like source, are used to detect the emission for each of the models, which are 
shown in Figure~\ref{fig:TheoreticalProfiles}.

To derive the strongest possible constraints, we compute upper limits on the 
integral flux above a given energy that accumulates results of several energy bins. 
The obtained limits depend on the assumed spectral shape.
%For both the \emph{semi-analytical} and \emph{extended} models,
We adopt the universal CR-proton spectrum found with hydrodynamical simulations \cite{2010MNRAS.409..449P},
that suggests a power-law momentum spectrum $p^{-\alpha}$ with $\alpha \approx 2.2$ at the 
energies of interest here.
%For the less specific \emph{isobaric} model the spectral index is free to vary and we assume a range of values, $2.1\leq\alpha\leq2.5$.
At very high energy, the $\gamma$-ray spectrum induced by pion decays should have 
approximately the same spectral index.
We convert this intrinsic spectrum into the spectrum
observed on Earth, taking into account the $\gamma$-ray absorption by the Extragalactic Background Light (EBL) \cite{Dominguez2011}
from the Perseus cluster at $z$=0.018.
Between 300\,GeV and 10\,TeV, the effect of the absorption can be approximated 
by an increase (softening) of the power-law index of about $0.13$ and a reduction of the 
differential flux at 1\,TeV of 17\%. Above 10\,TeV the absorption 
increases dramatically. A source with an intrinsic spectral index $\alpha=2.2$
would appear above 300\,GeV as a power-law with an index $\Gamma=2.33$ and a cutoff above 10\,TeV.
About 20\% of the flux above 1\,TeV and 60\% above 10\,TeV 
is absorbed during the travel to Earth. This effect was neglected in all previous papers,
which, therefore, overestimate the constrains induced by the flux upper limit. 

Table~\ref{tab:ULs} presents the 95\%-confidence-level upper limits (method from \cite{2005NIMPA.551..493R} with a total systematic uncertainty of 30\%)
of the integral flux between several energy thresholds $E_{th}$ and 10\,TeV for different models assuming a 
power-law spectrum with $\Gamma=2.33$. The upper limits are converted 
to the corresponding flux contained within the reference radius $0.15^\circ$ and 
the cluster virial radius\footnote{The cluster virial radius is here defined 
with respect to a density that is 200 times the critical density of the 
Universe.} $R_{200} \simeq 1.4^\circ$ \cite{2002ApJ...567..716R}. The 
conversion factors depend on the surface brightness distributions, considering 
also the signal contamination in the OFF region. 
%The latter is particularly important for 
%the \emph{extended} CR model, which have a quite flat surface brightness profile.
%For the \emph{point-like} assumption we estimated the background 
%from five OFF regions, while for the CR models we use a single region at 0.8$^\circ$ from the cluster centre.
%, as mentioned in Section~\ref{sec:2}.

%The effective area of MAGIC is relatively flat above 630\,GeV and the integral 
%flux upper limits do not depend strongly on the assumed spectral shape. 
%Therefore, the upper limits reported in Table~\ref{tab:ULs} are valid, within 2\%,
%for an observed spectral index range of $2.1<\Gamma<2.6$, which 
%corresponds to an intrinsic index range of about $2.0<\alpha<2.5$. We discuss the 
%implications of these results for the CR population in the Perseus cluster in 
%the next section.

\begin{table}[t!]
\caption{Integral flux upper limits within a radius of 0.15$^\circ$ and 1.4$^\circ$ from the Perseus cluster centre,
between $E_{th}$ and 10\,TeV for a power-law spectrum with an index $\Gamma=2.33$ [in unit of 10$^{-14}$\,cm$^{-2}$\,s$^{-1}$].} 
\begin{center}
%\resizebox{0.48\textwidth}{!}{
\begin{tabular}{lcccccc}
\hline\hline
%\phantom{\Big|}
%$E_{th}$ & model & ON\tablefootmark{a} & OFF\tablefootmark{a} & $\sigma_{LiMa}$\tablefootmark{b} & F$_{UL}^{0.15^\circ}$ & F$_{UL}^{1.4^\circ}$\tablefootmark{c}\\
$E_{th}$ & model & ON & OFF & $\sigma_{LiMa}$ & F$_{UL}^{0.15^\circ}$ & F$_{UL}^{1.4^\circ}$\\
\hline\\[-1.0em]
         &\emph{point-like}    & 332 & 304.1 & 1.4 & 12.2 &  12.2 \\
630\,GeV &\emph{isobaric}      & 1327 & 1256 &1.4 & 24.8 &  65.5 \\
         &\emph{semi-analytic} & 749 & 681 & 1.8 & 26.8 &  49.0 \\
         &\emph{extended}      & 1327 & 1256 & 1.4 & 25.6 &  124. \\
\hline
         &\emph{point-like}    & 159 & 157.5 & 0.1 & 3.84 &  3.84  \\
1.0\,TeV &\emph{isobaric}      & 675 & 652 & 0.6 & 10.7 &  28.3  \\
         &\emph{semi-analytic} & 369 & 352 & 0.6 & 9.77 &  17.9  \\
         &\emph{extended}      & 675 & 652  & 0.6 & 11.1 &  54.0  \\
\hline
         &\emph{point-like}    & 77 & 75.9 & 0.1 & 2.34 &  2.34  \\
1.6\,TeV &\emph{isobaric}      & 321 & 317 & 0.2 & 5.09 &  13.5  \\
         &\emph{semi-analytic} & 169 & 167 & 0.1 & 4.61 &  8.43  \\
         &\emph{extended}      & 321 & 317 & 0.2 & 5.26 &  25.6  \\
\hline
%         &\emph{point-like}    & 38 & 37.4 & 0.1 & 1.57 &  1.57  \\
%2.5\,TeV &\emph{isobaric}      & 143 & 153 & -0.6 & 2.18 &  5.75  \\
%         &\emph{semi-analytic} & 77 & 81 & -0.3 & 2.30 &  4.21  \\
%         &\emph{extended}      & 143 & 153 & -0.6 & 2.25 &  11.0  \\
%\hline
%2.5\,TeV &\emph{isobaric}      & - & - & - & 3.10 &  8.15  \\
%for zero &\emph{semi-analytic} & - & - & - & 2.81 &  5.15  \\
%excess   &\emph{extended}   & - & - & - & 3.21 &  15.5  \\
%\hline
\end{tabular}
%}
\end{center}
{\bf Notes.} 
ON and OFF are the number of events in the signal (ON) and background (OFF) regions, respectively.
$\sigma_{LiMa}$ is the significance of the excess expressed in standard deviations.
%For negative measured excess, we computed conservative upper limits assuming zero excess.
\label{tab:ULs}
\end{table}%

\section{Interpretation and discussion}
\label{sec:5}
One of the main uncertainties in modeling CR physics in clusters of galaxies is 
the CR spatial distribution, which is assessed by considering different spatial templates (\emph{isobaric}, \emph{semi-analytical} and \emph{extended}).
Another major uncertainty is the CR-acceleration efficiency, i.e., the fraction of the energy dissipated
at structure formation shocks which causes particle acceleration. 
Our \emph{semi-analytical} model is based on predictions by 
\cite{2010MNRAS.409..449P} derived from hydrodynamical simulations 
of clusters, which assume a maximum CR-proton acceleration efficiency 
$\zeta_\mathrm{p, max}= 50$\%. Here, we tie this 
to the parameter $A_\gamma$, just a multiplier for the whole formalism,  
that is $A_\gamma=1$ for $\zeta_\mathrm{p, max}= 50$\%. 
Smaller values of $A_\gamma$ correspond to smaller values of $\zeta_\mathrm{p, max}$.
%While the $A_\gamma$-$\zeta_\mathrm{p, max}$ dependence is generally non-linear, it is linear for energies $\gtrsim$\,10\,GeV.
%corresponding to pion-decay emission above $\gtrsim$\,1\,GeV.

\begin{figure*}[ht!]
\centering
\includegraphics[width=0.53\textwidth]{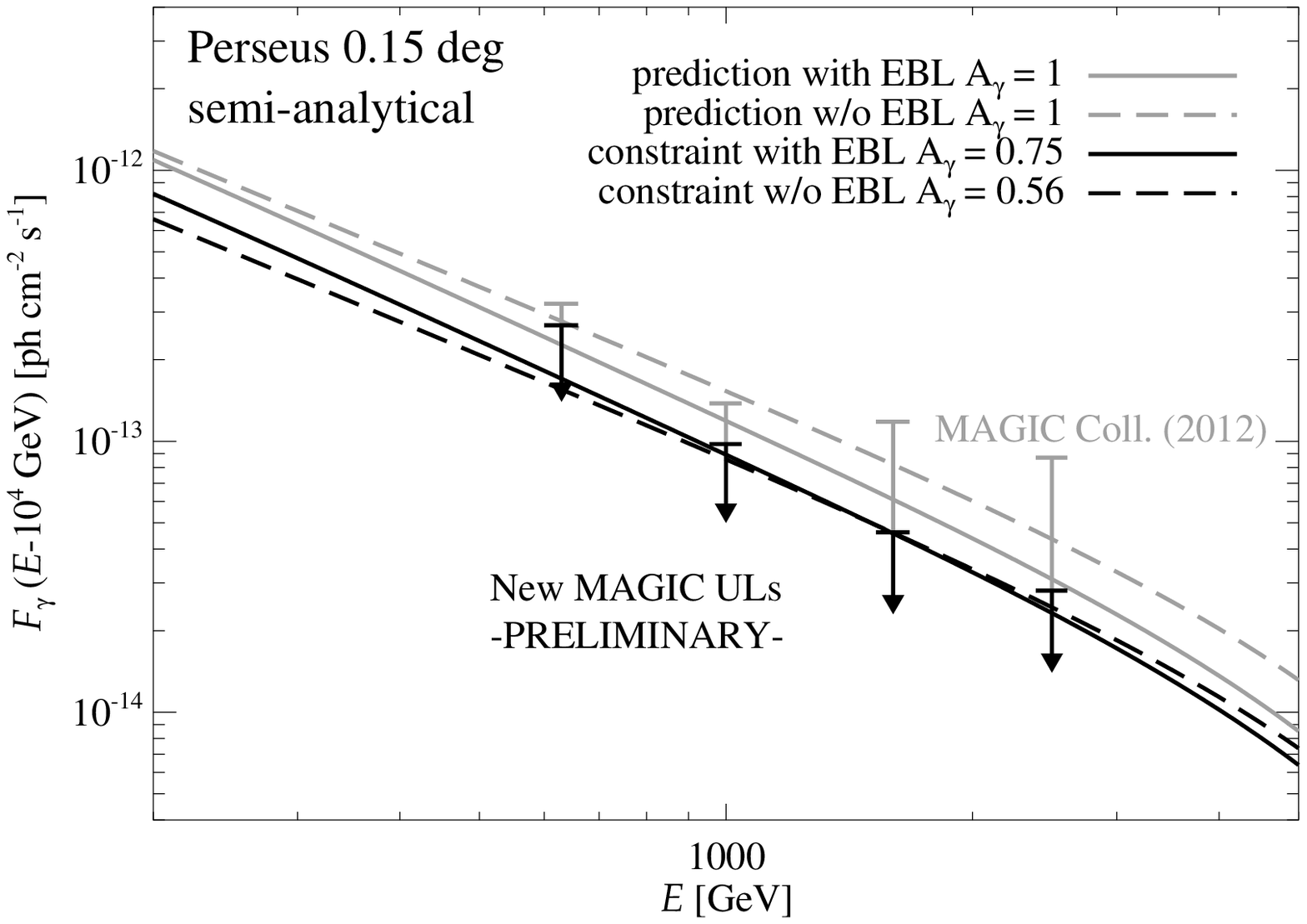}
\includegraphics[width=0.46\textwidth]{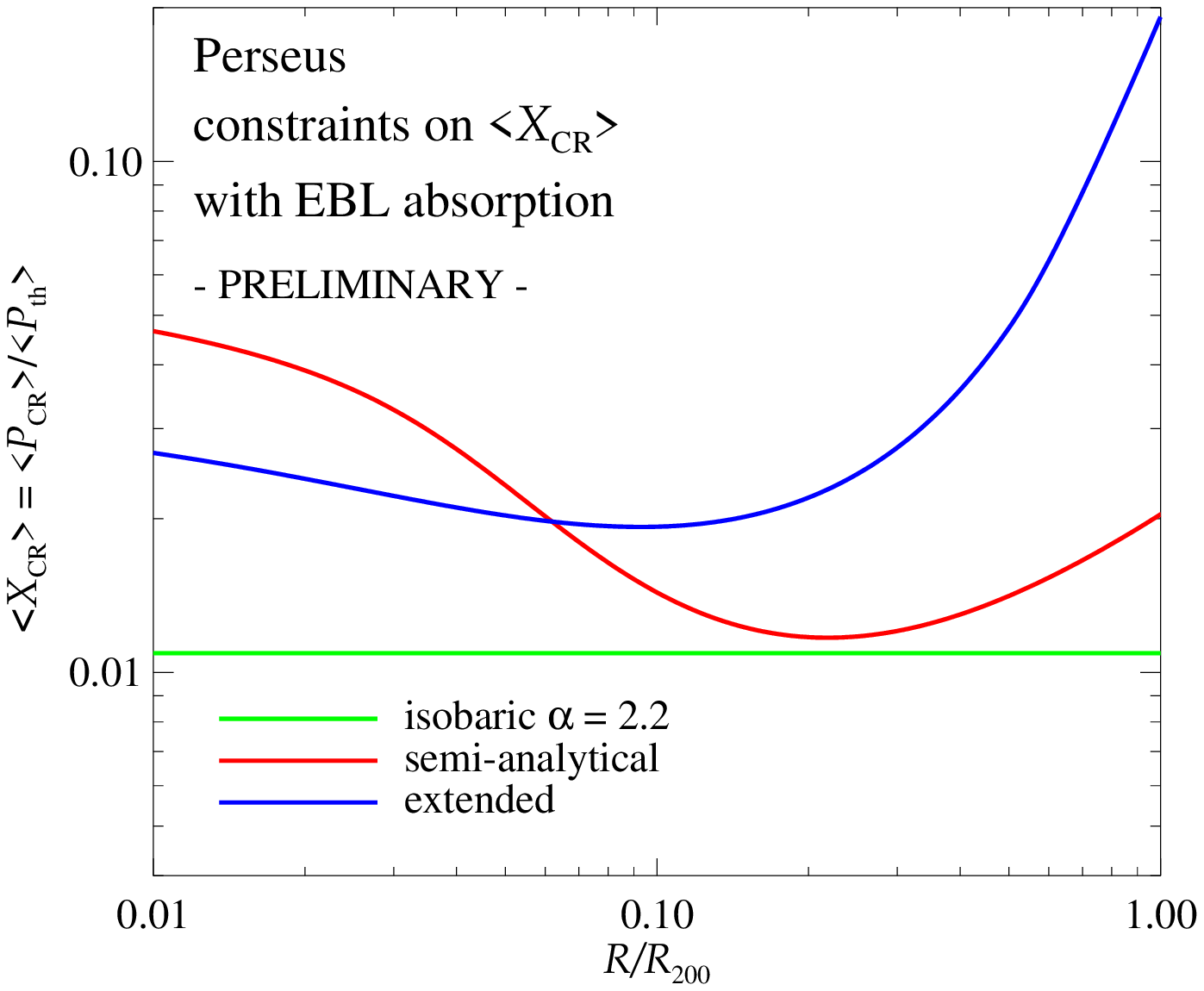}
\caption{{\bf \emph{Left.}} Integral flux upper limits and spectra within 
$0.15^\circ$ for the \emph{semi-analytical} model. We show the spectrum 
predicted by \cite{2010MNRAS.409..449P} and the ones adapted to match our upper limits.
%We also indicate the CR-multiplier $A_\gamma$ which is tied to the maximum 
%CR proton acceleration efficiency adopted in \cite{2010MNRAS.409..449P}; 
%$A_\gamma=1$ for a maximum acceleration efficiency of $\zeta_\mathrm{p, max}=50$\%. 
Shown in light gray are our previous flux upper limits from \cite{2011arXiv1111.5544M}.
{\bf \emph{Right.}} Volume-averaged $X_\mathrm{CR}$ within a given radius $R$, as constrained by the upper limits 
presented in this work for the \emph{semi-analytical}, \emph{extended}, and \emph{isobaric} models 
with $\alpha=2.2$.}
%We recall that in the \emph{isobaric} model $X_\mathrm{CR}(r)$ is constant by construction.
\label{fig:ConstrainedSpectra}
\end{figure*}

%The hadronic-induced emission scales linearly with $A_\gamma$, which fixes the 
%overall normalization of the CR distribution. 
The left-hand panel of Figure~\ref{fig:ConstrainedSpectra}, shows the integral $\gamma$-ray fluxes
predicted by \cite{2010MNRAS.409..449P} within $0.15^\circ$ from the centre, $A_\gamma=1$,
and the highest fluxes allowed by our observational upper limits. 
Compared to \cite{2011arXiv1111.5544M}, we accumulate three times more data and derive upper limits that are significantly lower.
$A_\gamma$ is constrained to be $\leq 0.56$ when neglecting the EBL absorption and $\leq 0.75$ with the EBL\cite{Dominguez2011}.
This corresponds to a maximum CR-proton acceleration efficiency of 28\% and 37\%, respectively.
The latter should be considered the reference value while the one without EBL absorption is given
to allow comparison to previous results.

The constraints on the volume-averaged CR-to-thermal pressure profiles  
for the three CR-distribution models are shown in Figure~\ref{fig:ConstrainedSpectra}-left.
$<X_\mathrm{CR}>$ results to be 
below 1--2\% within $0.15^\circ$ ($\approx 0.11 \times R_{200}$).
When considering the full galaxy cluster volume up to $R_{200}$, 
$<X_\mathrm{CR}>$ is below 2\% for the \emph{isobaric} and \emph{semi-analytical} models, but
significantly less constrained, below 19\% for the \emph{extended} model. This latter 
weak constraint was expected because the CR distribution is significantly flatter in this case.
%building up to large $X_\mathrm{CR}$ values in the cluster outskirts where the thermal pressure becomes very small.
%as the ICM density decreases.

The Perseus cluster hosts a bright radio mini-halo. 
Assuming the hadronic model of radio halos, i.e., synchrotron emission generated by 
secondary electrons produced in hadronic interactions between the CR and ICM,
the pion-decay $\gamma$-ray emission is directly linked to radio signal.
The intensity of the synchrotron emission depends on the amount of electrons, 
proportional to the hadronically-induced $\gamma$ rays, and the local magnetic field.
This emission can be reduced due to the electron energy lost through other channels, mainly inverse Compton.
In the regime of very strong magnetic fields, the electrons lose all their energy through synchrotron and
we can derive the expected $\gamma$-ray emission. For assumptions of CR spectral index $\alpha \leq 2.1$,
the expected $\gamma$-ray emission is above our upper limits. For softer CR spectrum, our $\gamma$-ray upper limits
can be turned into lower limits on the magnetic field. 
This constraint, depending on the photon field considered for the inverse Compton scattering, is still under study.
 
\textbf{Acknowledgment}\\
\small{We would like to thank
the Instituto de Astrof\'{\i}sica de Canarias
for the excellent working conditions
at the Observatorio del Roque de los Muchachos in La Palma.
The financial support of the German BMBF and MPG,
the Italian INFN and INAF,
the Swiss National Fund SNF,
the ERDF under the Spanish MINECO (FPA2012-39502), and
the Japanese JSPS and MEXT
is gratefully acknowledged.
This work was also supported
by the Centro de Excelencia Severo Ochoa SEV-2012-0234, CPAN CSD2007-00042, and MultiDark CSD2009-00064 projects of the Spanish Consolider-Ingenio 2010 programme,
by grant 268740 of the Academy of Finland,
by the Croatian Science Foundation (HrZZ) Project 09/176 and the University of Rijeka Project 13.12.1.3.02,
by the DFG Collaborative Research Centers SFB823/C4 and SFB876/C3,
and by the Polish MNiSzW grant 745/N-HESS-MAGIC/2010/0.}

\end{document}